\documentclass[journal]{IEEEtran}
\usepackage{cite}

\ifCLASSINFOpdf
  \usepackage[pdftex]{graphicx}
\else
\fi
\usepackage{amsmath}
\usepackage{tikz}
\definecolor{icefill}{HTML}{D8EAF5}
\definecolor{coastline}{HTML}{8FA6AE}

\ifCLASSOPTIONcompsoc
  \usepackage[caption=false,font=normalsize,labelfont=sf,textfont=sf]{subfig}
\else
  \usepackage[caption=false,font=footnotesize]{subfig}
\fi

\usepackage{macros}

\begin{document}
%
\title{Bayesian 
Retrieval of
L-Band Absorption by Ice using SMOS Brightness Temperatures over the Antarctic Plateau}
%
%
%

\author{Hippolyte~Signargout,
Ghislain~Picard,
Olivier~Zahm,
Ange~Haddjeri,
Marion~Leduc-Leballeur,
Pierre~Zeiger,
Marco~Brogioni,
and~Giovanni~Macelloni%
\thanks{Hippolyte Signargout, Ghislain Picard, Ange Haddjeri and Pierre Zeiger are with the Institut de Géosciences de l'Environnement, Centre National de Recherche Scientifique, Grenoble, France. e-mail: hippolyte.signargout@univ-grenoble-alpes.fr}
\thanks{Olivier Zahm is with Inria Grenoble Alpes, Grenoble, France.}
\thanks{Marion Leduc-Leballeur, Marco Brogioni, and Giovanni Macelloni are with the Institute of Applied Physics “Nello Carrara” in National Research Council, 50019 Sesto Fiorentino, Italy.}%
}

%
%

\markboth{IEEE Transactions on Geoscience and remote sensing,~Vol.~?, No.~?, ?~?}{}%
%

\maketitle

\begin{abstract}
The low absorption of microwaves at L-band in ice enables the retrieval of internal ice sheet parameters from satellite observations. 
However, knowing the precise value of the ice absorption $\ka(T)$ as a function of temperature is a strict prerequisite.
The existing knowledge is well constrained above 2 GHz and under 200 MHz, lacking constraints in the P- and L-bands.
In this study, we infer ice absorption values using the spatial variations of L-band radiometric observations across Antarctica and comparing them with internal ice temperature variations through a simple emission model. More importantly, we adopt a probabilistic approach to determine a credible interval for absorption, using a Bayesian retrieval method. 
Our algorithm seeks the optimal multiplier factor $\alpha$ to apply to the current lowest estimate ice absorption formulation $\ka(T)$ proposed by Mätzler 2006, and computes the probability distribution of $\alpha$. 
The results show the absorption is most likely 1.5 to 1.7 times higher than this lowest formulation, which reduces the actual sensing depth at L-band. 
\end{abstract}


%
\IEEEpeerreviewmaketitle

\section{Introduction}

\IEEEPARstart{T}{he} electromagnetic absorption coefficient of ice, driven by the imaginary part of its complex permittivity, controls the emission and penetration depths of microwave signals in ice sheets. It is a critical parameter of ice sheet emission models, which underpin the retrieval of snow and ice physical properties in the ice sheet dry zones from satellite observations. While the K and Ka bands (18--40 GHz) have been extensively used to study near-surface properties (0-10\,m) \cite{zwally77,schneider04,picard06,brucker10}, lower frequencies are less absorbed and the Soil Moisture and Ocean Salinity (SMOS) and Soil Moisture Active and Passive (SMAP) missions \cite{kerr10,entekhabi10}, operating at 1.41 GHz (L-band), have been shown to be sensitive to internal parameters such as the ice sheet internal temperature profile (100 to 1000\,m deep) \cite{macelloni16, lv25}. Owing to their essential contribution for climate monitoring \cite{zhou26}, the continuity of L-band measurements is guaranteed with the upcoming European Space Agency's (ESA) Copernicus Imaging Microwave Radiometer (CIMR) \cite{donlon23}. Furthermore, the CryoRad candidate mission to ESA 12th Earth Explorer (EE12) proposes continuous L- and P-band spectral measurements (0.4--2\,GHz) for the first time, with the aim to retrieve deep ice temperature and basal melted or frozen state \cite{macelloni26, cryorad26}. Although it is known that the absorption of ice has a minimum in L- and P-band, its precise value remains poorly constrained despite being a critical parameter to evaluate the performance of  deep ice sheet applications.  

The low absorption of ice at L-Band, implying \(e\)-folding depth in hundreds to thousands of meters, makes laboratory measurements technically difficult, resulting in a large spread among estimated values.
Two widely used formulations for absorption are derived from the permittivity models of Tiuri \textit{et al} \cite{tiuri84} and Mätzler \textit{et al} \cite{matzler06} (T84 and M06 hereinafter). They are based on measurements conducted at different frequencies and extrapolated to 1.41 GHz for temperatures values typical of the Antarctic ice sheet. As a consequence, the 1.41 GHz absorption values obtained from T84 and M06 differ by a factor of 4 to 8 depending on the temperature.
Other measurements \cite{vickers75,warren84,matsuoka96} fall in between or higher, without resulting in a consensual value.

Spaceborne L-band measurements can be used as a tool to help constrain the absorption values among the different estimates.
Macelloni \etal \cite{macelloni16} modeled the L-band \TB variations over three transects in Antarctica using both T84 and M06 formulations.
Since spatial variations of the internal ice temperature are greater at the surface than at the bottom of the ice sheet, the absorption value influences the spatial variations of the modeled \TB.
The comparison between modeled and observed SMOS \TB therefore gives a constraint on which absorption value is likely. 
Over the three transects, this comparison shows that using M06 tends to produce slightly larger variations, whereas T84 produces weaker variations, indicating that the absorption is likely between M06 and T84.
However, deriving precise absorption values by matching modeled and observed measurements across the ice sheet is challenging as  \TB depends both on deep ice emissions, constrained by the ice sheet internal temperature profile and the absorption, and on reflections occurring near the surface in the firn \cite{leduc15, tan21}.

An attempt to disentangle these two effects in order to estimate more precise absorption values was performed by
Passalacqua \textit{et al} \cite{passalacqua2018} (P18 hereinafter).
It compares SMOS \TB over a large area of Antarctica with \TB modeled using a simple emission model, and seeks to minimize their difference by varying the absorption.
Their emission model uses a single multiplicative transmission factor to represent the firn layering reflections. This additional parameter makes the minimization underdetermined.
To regularize the minimization problem, the spatial variations of this factor are assumed  uncorrelated to the output of the emission model. Its resolution allows retrieval of absorption (for a range of temperatures) along with the transmission factor for each pixel.
The results confirm M06. 
However, their retrieval framework does not quantify the uncertainty of the retrieved values.
Moreover, a few methodological choices can be improved. All observations were given equal importance, ignoring any spatial correlation in the model error. The lack of spatial correlation between \TB and the transmission factor, while central to the method, is also debatable. Another limitation arises from neglecting the absorption temperature dependence over ice sheet internal temperature profiles. To limit the impact of this approximation, their method was applied to chunks of ice-sheet with similar mean temperature. Finally, the knowledge on internal temperature profiles in Antarctica has improved since the study thanks to recent improvements on geothermal heat flow \cite{stal2021} and bedrock elevation \cite{pritchard2025}.

This paper aims at retrieving L-band ice absorption by matching the spatial variations of observed and modeled \TB across Antarctica. 
Importantly, the retrieval method aims at providing rigorous estimates of the uncertainties associated with the estimated absorption using Bayesian inference for inverse uncertainty propagation. Considering a reference absorption formulation \(\kao(T)\), we compute the probability density function (\pdf) of a scaling factor \alp on \kao at 1.41 GHz using radiometric measurements from the SMOS satellite, assuming that the temperature dependence of \kao is correct. M06 is used as the reference here.
The emission model used represents the reflection effect of snow layering in the firn\cite{leduc15,tan15} with a single transmittance variable \tra, considered constant in zones where we expect low variations.
The use of Bayesian inference also permits the inference of \tra along with statistics on the model error (its spatial correlation structure and its standard deviation). 
The retrieval problem is partly solved analytically before using brute force evaluation on the remaining 2-dimensional problem.

The paper is organized as follows: \Cref{sec:mem} details the inference method, models and data; \cref{sec:res} presents the results and verifies \textit{a posteriori} the underlying hypotheses ; \cref{sec:dis} places our results in the context of previous estimates.

\section{Data and Method}
\label{sec:mem}

\subsection{Data}
\label{ssc:mat}
The method uses the MIRAS L-Band (1.41 GHz) radiometer observations from the SMOS satellite \cite{kerr10} at vertical polarization (unless specified). While SMOS interferometric measurements is multi-angular, we only select the 52.5° incidence angle, close to Brewster angle to reduce reflections in the firn at vertical polarization.
In this study, the SMOS dataset is an \textit{enhanced resolution} product posted on a 12.5 km polar stereographic grid, with an effective spatial resolution previously evaluated to 45 km \cite{zeiger2024}.  In each pixel, the dataset is averaged over ten years (2012 - 2022) following \cite{macelloni16}. We also mask out areas with insufficient observations (latitude lower than -80°) and those consistently subject to melt \cite{picard06}.

Several geophysical parameters are needed as input of the thermal model feeding the emission model. Ice sheet thicknesses (ice, firn and snow total) were taken from Bedmap3 \cite{pritchard2025}, geothermal heat flux from Stål et al. 2021 \cite{stal2021}, mean yearly snow accumulation from Agosta et al., 2019 \cite{agosta2019}, and mean yearly surface temperature from Freville et al., 2014 \cite{freville2014}. In addition, surface ice velocity \cite{mouginot2019} were used to mask out fast-flow regions in which our models are invalid.

\subsection{Method}
\subsubsection{Emission modeling}
\label{sss:mod}

The modeling of L-band emission from the ice sheet is performed in two steps. First, we compute internal temperature profiles across the valid area using the Robin model and geophysical parameters as inputs. Second, we compute in each pixel the modeled \TB (denoted \(\tb^m\)) using a simple emission model.

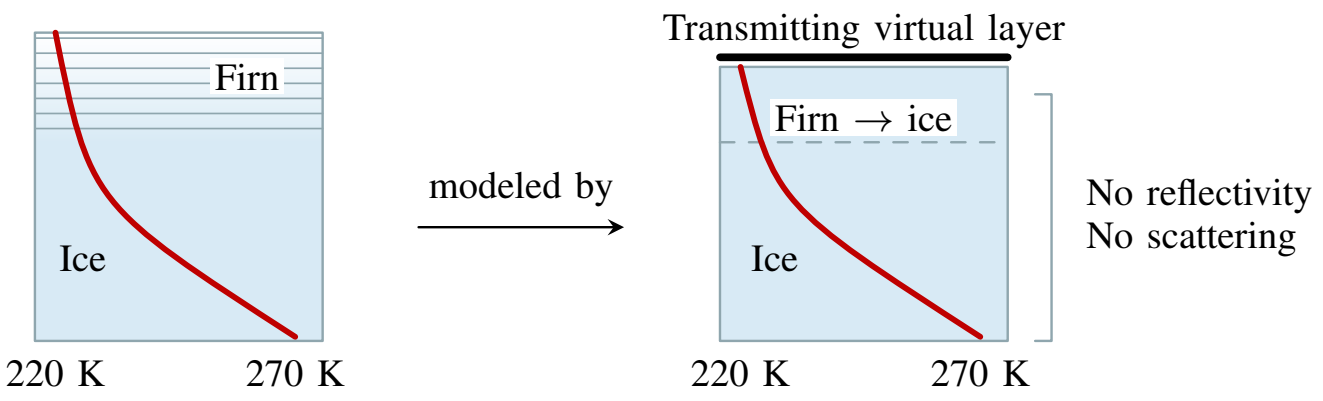
\begin{figure}[htb]
    \centering
    \resizebox{\columnwidth}{!}{%
    \begin{tikzpicture}[font=\footnotesize, line join=round, line cap=round]
      \fill[icefill] (0.4,0.65) rectangle (2.5,2.1);
      \shade[top color=white, bottom color=icefill] (0.4,2.1) rectangle (2.5,2.9);
      \draw[coastline, line width=0.5pt] (0.4,0.65) rectangle (2.5,2.9);
      \foreach \y in {2.2,2.31,2.42,2.53,2.64,2.75,2.86}
        \draw[coastline, line width=0.35pt] (0.4,\y) -- (2.5,\y);
      \node[fill=white, inner sep=1pt] at (1.95,2.58) {Firn};
      \node at (0.75,1.25) {Ice};

      \fill[icefill] (5.4,0.65) rectangle (7.5,2.65);
      \draw[coastline, line width=0.5pt] (5.4,0.65) rectangle (7.5,2.65);
      \draw[coastline, dashed, line width=0.5pt] (5.4,2.1) -- (7.5,2.1);
      \draw[black, line width=1.6pt] (5.4,2.72) -- (7.5,2.72);
      \node[fill=white, inner sep=1pt] at (6.45,2.28) {Firn $\to$ ice};
      \node at (5.8,1.25) {Ice};
      \node at (6.45,2.91) {Transmitting virtual layer};

      \node at (3.95,1.76) {modeled by};
      \draw[-{stealth}, line width=0.5pt] (3.2,1.48) -- (4.7,1.48);

      \draw[red!75!black, very thick]
        (0.55,2.9) .. controls (0.8,1.65) .. (2.3,0.68);
      \draw[red!75!black, very thick]
        (5.55,2.65) .. controls (5.8,1.65)   .. (7.3,0.68);

      \node[anchor=north] at (0.55,0.64) {220 K};
      \node[anchor=north] at (2.3,0.64) {270 K};
      \node[anchor=north] at (5.55,0.64) {220 K};
      \node[anchor=north] at (7.3,0.64) {270 K};

      \draw[coastline, line width=0.5pt]
        (7.7,0.65) -- (7.82,0.65) -- (7.82,2.45) -- (7.7,2.45);
      \node[align=left, anchor=west] at (7.95,1.55)
        {No reflectivity\\No scattering};
    \end{tikzpicture}%
    }
    \caption{Typical real ice sheet profile with low-density firn (left) and its ice-equivalent representation (right). The firn is condensed into ice and the layers are replaced by a single transmitting-only layer. Internal temperature profiles are shown in red.}
    \label{fig:scm}
\end{figure}

\paragraph{Temperature profiles}
The internal temperature profile is calculated with the one-dimensional Robin model \cite{robin55}.
It takes as input for one pixel the ice thickness, geothermal heat flux, mean yearly snow accumulation and mean yearly surface temperature, and outputs the internal temperature profile $T(z)$. 
We rectify the ice sheet thickness by condensing snow layers into pure ice layers of equivalent mass (\cref{fig:scm}). The average density of snow layers is given by the Herron and Langway model \cite{herron80} as formulated in \cite{arthern2010}. 

The Robin model neglects the horizontal ice advection, and can reasonably be applied where ice velocity is lower than 5~m~yr\(^{-1}\) \cite{macelloni19}, which limits our study to locations where this constraint is met in surface ice velocity observations \cite{mouginot2019} (areas represented in \cref{fig:map}). 
In the resulting study area, melt occurs only in 0.01\% of the pixels and at most once every 3 years \cite{picard06}, allowing to assume dry-snow conditions.

\paragraph{Brightness temperatures}

We adopt a simple emission model to compute the brightness temperature from the internal temperature profile. We consider that absorption dominates, and neglect volume scattering in snow and ice because snow grain size is much lower than the wavelength \cite{zwally77,matzler87}. Setting aside internal reflections for the moment, the emission of the ice column at the surface is
\begin{equation}
  \tb^{\text{ice}} = \int_{0}^{d}
  {T(z)}\;{\frac{\kappa_a\left({z}\right)}{\mu}}
  \;{e^{-\displaystyle\int_{0}^{z}\frac{\kappa_a(z')}\mu\,dz'}}\,dz
\end{equation}
where \(d\) is the total ice equivalent thickness, and \(\mu = \cos \phi\) the cosine of the refraction angle (at incidence 52.5°).
\(T(z)\) and \(\ka(z)\) are the temperature and absorption of ice at depth \(z\). The ice column is discretized into 30 layers, as tests with more layers did not change the output \TB (\(<0.001\)\xspace K for 250 layers). 
Each layer has a linear temperature profile, and the temperature at the interfaces is taken from the Robin's model calculation. The absorption in each layer is calculated by 
\begin{equation}
    \ka(z) = \alp \; \ka^0\paren{f=1.41\;\text{GHz}, T = \overline{T^{\text{lay(z)}}}}
\end{equation}
where \kao is a reference absorption formulation which depends on frequency \(f\) and temperature \(T\). Unless specified, the formulation used is M06 given by Mätzler  \cite{matzler06}. The parameter \alp is the unknown scaling factor of the absorption that we aim to estimate as precisely as possible. \(\overline{T^{\text{lay}(z)}}\) is the mean temperature in the layer at depth $z$. We add the contribution of the bedrock to get the total \textit{effective temperature} at the surface.
\begin{align}
  \quad T_E &={\tb^{\text{ice}} + \tb^{\text{bed}}\;e^{-\displaystyle\int_{0}^{d}\frac{\kappa_a(z)}\mu\,dz}}
\end{align}

We consider the bedrock totally absorbent with a temperature equal to that of the ice bottom as ice velocity is low: \(\tb^{\text{bed}} = T(d)\).

Finally, although we selected observations at vertical polarization and near Brewster angle, the multiple reflections in the firn have a small yet noticeable effect on \TB. Advanced modeling studies  highlighting this effect \cite{leduc15,tan21, bai23} rely on input firn density profiles with a very high vertical sampling resolution (3-10 cm), down to at least 100 m. As this information is not available at continental scale, we take into account the effect of the reflecting firn layers in a simplified way, as a single total transmittance factor \(\tra\) (\cref{fig:scm}), to compute the modeled brightness temperature $T_B^m$: 
\begin{equation}
\label{eqn:ttt}
  \tb^m = \tra \; T_E 
\end{equation}
This factor is unknown in every point, and is inferred along with the other model parameters during the Bayesian inference.

\paragraph{Resampling of the modeled \TB}
\label{par:smo}
The variations of the observed \TB are smoothened by the satellite footprint being larger than the grid resolution of the product used in this study \cite{zeiger2024}.
 In order to reproduce the footprint and match comparable \TB variations, the modeled \TB are computed on a 4 km grid using \cref{eqn:ttt} and are convolved with a Gaussian kernel of half intensity diameter 45 km \cite{zeiger2024}. When the convolution crosses points outside of the study area, the result is normalized by the sum of the weights of the included pixels in the kernel. If the normalization induces more than 3\% change, the location is itself excluded.
 
\subsubsection{Bayesian Inference}
\label{sss:bys}

We compare the modeled \(\tb^m\) with observations denoted \(\tb^o\) through the random variable 
\begin{equation}
    \err = {\tb^o} - {\tb^m}.
\end{equation} 
Bayesian inference computes  the desired \pdf, called \textit{posterior probability}, as the product of the \textit{likelihood} of the parameter set given the observations and the \textit{prior distribution} of the parameters, which are the absorption factor \alp, transmittance \tra and hyper-parameters \(\sigma\) and \lda introduced in the following paragraphs:
\begin{equation}
  \p(\alpha, \tra, \sigma, \lda \mid \bm{\tb^o}) \propto \p(\bm{\tb^o} \mid \alpha, \tra, \sigma, \lda)  \p(\alpha, \tra, \sigma, \lda)
\end{equation}
Bold letters represent vectors of \TB, indexed here by their location.

\paragraph{Likelihood}
The likelihood follows the distribution of the error \(\bm\err\), which comprises multiple sources of uncertainty including observation uncertainty, model uncertainty, and input geophysical parameter uncertainties. We assume all sources of error to result in unbiased normal random variables. Their sum  \(\bm\err\) is then an unbiased normal law.
All sources of error are correlated for nearby pixels and this correlation is expected to decrease with the distance.
We set the covariance matrix of \(\err\) as \(\sigma^2\Sigma(\lda))\) where \(\sigma\) is a variable standard deviation and \(\Sigma(\lda)\) is a unit-diagonal Matérn 3/2 kernel of length-scale \lda\xspace:
 \begin{equation}\label{eq:sigma}
   \Sigma(\lda)_{ij} = \paren{1 + \sqrt3\frac{{d_{ij}}}{\lda}}{e^{-\sqrt3\frac{{d_{ij}}}{\lda}}}, \quad i,j\in\intset{1}{n}
 \end{equation}
 where \(d_{ij}\) is the geographical distance between points \(i\) and \(j\) and \(n\) is the number of pixels.
We have
\begin{equation}
    \bm\err\sim\mathcal{N}(0, \sigma^2\Sigma(\lda))
\end{equation}
where the standard deviation \(\sigma\) and the length-scale \lda, too complex to estimate, are set as hyper-parameters to infer.
Finally, the likelihood writes:

\begin{equation}
  P(\bm{\tb^{o}} \mid \alpha, \tra, \sigma, \lda) = \frac{e^{-\frac12\norm{\bm{\tb^{o}} - \bm{\tb^{m}}(\alpha, \tra)}_{\inv{\paren{\sigma^2\Sigma(\lda)}}}^2}}{\sigma^n\sqrt{2\pi|\Sigma(\lda)|}}
  \label{eq:lkh}
\end{equation}

 \paragraph{Prior distributions} 

The prior distributions represent the expert knowledge on the parameters before using the observations. Here, we do not have objective prior knowledge and want to let the observations fully constrain the results.
To this end, independent non-informative one-dimensional Jeffrey's priors \cite[S2.8]{gelman95} are selected for \(\alp,\, \tra,\, \sigma~\text{and}~\lda\). Compared with other possible non-informative priors
such as uniform priors, Jeffreys priors have the remarkable
property of being invariant under transformations of the
parameters being estimated.
This principle is achieved by setting high probabilities to values which have a high impact on the likelihood, resulting in the following definition:
\begin{equation}
\label{eqn:jef}
  \p(X) \propto \sqrt{ E_X\paren{\paren{\frac{\partial\log P(\bm{\tb^{o}} \mid X)}{\partial X}}^2}}, \quad X\in \{\alp, \tra, \sigma, \lda\}.
\end{equation}
Note that we avoid the difficulty of a multi-dimensional Jeffrey's prior by setting a nominal value \(\overline Y\) for each other parameter \(Y\) when computing the prior for \(X\). 
\Cref{eqn:jef} gives \(\p(\tra) = 1\) and \(\p(\sigma) = 1/\sigma\). The priors \(\p(\alpha) = \norm{\frac{\partial \bm{\tb^m}(\alpha, \overline t)}{\partial \alpha}}_2\) and \(\p(\lda)\) are computed numerically and show little dependence to the chosen nominal values.
These priors are improper: they do not integrate to 1 over the parameter space and can thus not be considered as probability distributions on their own. However, we show in \cref{sss:imp} that once multiplied by the likelihood, the posterior is integrable. 

\subsubsection{Implementation}
\label{sss:imp}

\begin{figure}[!hbt]
  \includegraphics[width=\linewidth]{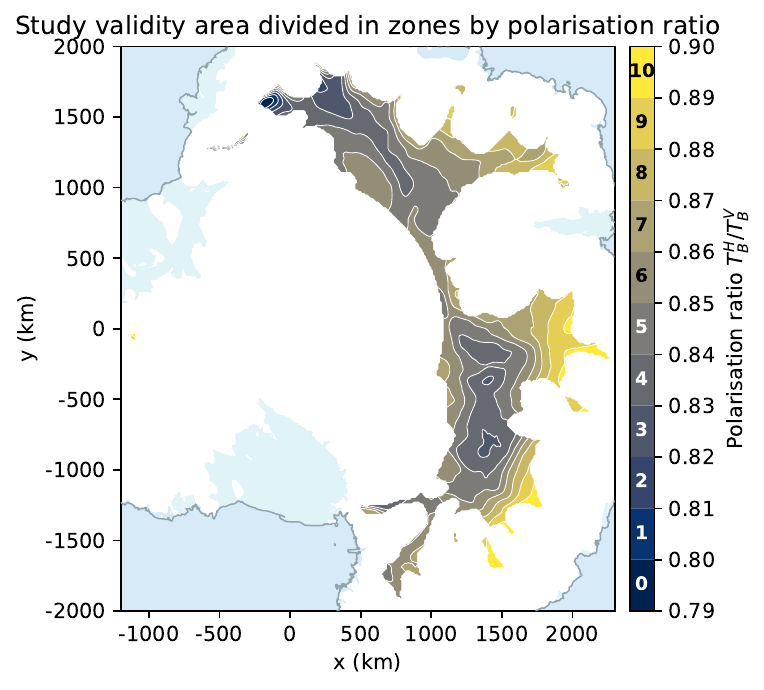}
\caption{Polarization ratio \(\tb^H/\tb^V\) from SMOS observations over the study area discretized in 11 zones in which the firn transmittance \(\tra\) is considered constant.}\label{fig:map}
\end{figure}

\paragraph{Layering zones}
\label{par:lay}
The transmittance at vertical polarization is affected by firn layering reflections. These reflections also greatly affect the polarization ratio \(\tb^H/\tb^V\) that can serve as a proxy. To constrain $\tra$, we 
partition Antarctica
into \(K = 11\) zones with nearly constant polarization ratios,
obtained by geometrically partitioning the range of observed
ratios (\cref{fig:map})
and assume $\tra$ is constant across each zone.
The polarization ratio is computed at an incidence angle
of 52.5° and constitutes the only use of horizontally polarized
observations \(\tb^H\) in this study; the vertically polarized data \(\tb^V\)
are the same as those used throughout the rest of the analysis.

The precision of the model is expected to vary between zones, which means the amplitude of the errors may differ. We infer one transmittance \tra and one standard deviation \(\sigma\) per zone.
\begin{align}
    \bm t &= \paren{\tra_0,\dots,\tra_{K-1}}\\
    \bm\sigma &= \paren{\sigma_0,\dots,\sigma_{K-1}}
\end{align}

\paragraph{Factorization of the posterior computation}
\label{par:mar}
The complexity of computing the posterior distribution, and thus the achievable precision for a given computation time, explodes with the dimension of the parameter space. We here divide the dimension of the problem by two by computing first \(\p(\alpha,  \lda \mid \bm{T_{B}^o})\). We then use this distribution to draw a sample of \((\alpha, \lda)\) and finally sample \(\bm t\) and \(\bm\sigma\) from \(\p(\bm t,  \bm\sigma\mid \alpha, \lda, \bm{T_{B}^o})\):
\begin{equation}
  \p(  \alpha, \bm t, \bm\sigma, \lda | \bm{\tb^o}) = \p(\alpha,  \lda | \bm{\tb^o})  \p(\bm\sigma | \alpha,  \lda, \bm{\tb^o}) \p(\bm t | \bm\sigma, \alpha,  \lda, \bm{\tb^o})
\end{equation}
We use an additional hypothesis to recover an analytical formulation for \(\p(\alp, \lda\mid\bm{\tb^o})\): \(\bm{\err_k}\)  errors are independent from each other.
Mathematically speaking, this means \(\Sigma\) is block-diagonal with observations ordered by zone.
\begin{equation}
    \Sigma = \text{diag}\paren{\Sigma_0,\dots,\Sigma_{K-1}}
\end{equation}
As the \(\tra_k\) and \(\sigma_k\) variables only affect \(\bm{\eta_k}\) we can factorize the likelihood.
\begin{align}
    \p(\bm{\tb^o}\mid\alp, \bm t, \bm\sigma, \lda) &= \prod\limits_0^{K - 1}\p(\bm{\tb^o}\mid\alp, \tra_k, \sigma_k, \lda)\\
    &= \prod\limits_0^{K - 1} 
    \frac{e^{-\frac12 \norm{\bm{T_{Bk}^o} - \tra_k\bm {T_{Ek}}(\alpha)}_{\inv{\paren{\sigma_k^2\Sigma_k(\lda)}}}}}
    {\sigma_k^{n_k}\sqrt{2\pi\left|\Sigma_k(\lda)\right|}}\label{eqn:ind}
\end{align}
with \(T_E(\alp)\) the effective temperature modeled with \alp and the \(k\) subscripts indicate the quantity is taken only from zone~\(k\).
Marginalization over \(\bm t\) and \(\bm\sigma\) is now possible analytically (details in Appendix \ref{app:mar}), yielding:
\begin{equation}
  \label{eqn:ppp}
  \p(\alpha,  \lda|\bm{{\tb^{o}}}) \propto \frac{\p(\alpha, \lda)}{\sqrt{|\Sigma(\lda)|}}
  \prod\limits_{k = 0}^{K-1}
\frac{\| \bm{T_{Bk}^o}- \tra_{k}^{\alpha} \bm{T_{Ek}}(\alpha)\|_{\Sigma_k(l)^{-1}}^{1-n}}{\sqrt{\norm{\bm{T_{Ek}}(\alp)}_{\inv{\Sigma_k(\lda)}}}}
\end{equation}
where
\begin{equation}
t_k^\alpha=\tfrac{\trsp{\bm{T_{Ek}}(\alp)}\inv{\Sigma_k(\lda)}\bm{T_{Bk}^o}}{\norm{\bm {T_{Ek}}(\alp)}_{\inv{\Sigma_k(\lda)}}}
\end{equation}
is the minimizer of \(\tra \mapsto \| \bm{T_{Bk}^o}- \tra \bm{T_{Ek}}(\alpha)\|_{\Sigma_k(l)^{-1}}\).
The following inverse gamma and normal distributions are then used for sampling (proof in Appendix \ref{app:spl}):
\begin{align}
  \sigma_k^2 &\mid \alpha,\lda,\bm{{\tb^{o}}} \sim \mathcal{IG}\left(\frac{n-1}{2},\ \frac{\| \bm{T_{Bk}^o}- \tra_k^\alpha \bm{T_{Ek}}(\alpha)\|_{\Sigma_k(l)^{-1}}}{2}\right)\\
\tra_k &\mid \sigma_k,\alpha,\lda,\bm{{\tb^{o}}}\sim \mathcal N\left(t_{k}^{\alpha},\frac{\sigma_k^2}{\norm{\bm{T_{Ek}}(\alp)}_{\inv{\Sigma_k(\lda)}}}\right)
\end{align}
\paragraph{Probability computation} 
The posterior distribution \(\p(\alpha, \lda \mid \bm{{\tb^{o}}})\) is computed on a 300 $\times$ 200 grid covering \(0.7\leq\alp\leq4\) and \(2.5\leq\lda\leq250\), first with \cref{eqn:ppp} and then normalized to a valid \pdf.
Samples for \alp and \lda are drawn from this discrete weighted distribution.
The \pdf of \(\tra\) and \(\sigma\) are classical and samples are drawn without requiring a particular algorithm.
\paragraph{Downsampling}

The retrieval of parameters \(\sigma\) and \lda for a Matérn kernel with a large number of pixels inside the support of the kernel is a known complex problem \cite{loh23}. To limit this difficulty, we reduce the amount
of nearby pixels by sampling one every other pixels in each direction, that is every 25 km. This is still much less than the effective resolution of SMOS \TB ($\sim$45~km) which avoids loss of information.

\section{Results}
\label{sec:res}

We first detail the effect of \(\alpha\) on the correlation between the modeled and observed \TB (\cref{ssc:vis}), before discussing the resulting posterior distributions (\cref{ssc:pst}) and finally post-evaluating our initial hypotheses (\cref{ssc:hyp}).

\subsection{Emission Model Sensitivity to Ice Absorption}
\label{ssc:vis}
\begin{figure}[!bht]
\includegraphics[width=\columnwidth]{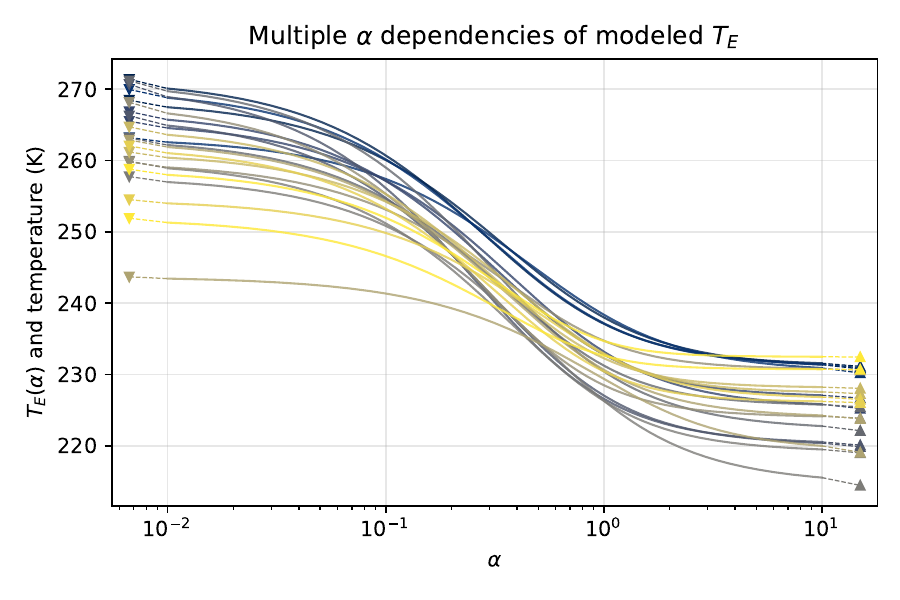}
\caption{Modeled effective temperature with respect to \alp (log-scale) for  two random pixels per zone. Colors refer to different zones (\textit{cf.} Figure~\ref{fig:map}). The bedrock temperature is indicated on the left (\(\bigtriangledown\)) and the surface temperature on the right (\(\bigtriangleup\)).}
\label{fig:alp}
\end{figure}

The effect of the ice absorption parameter \alp on the modeled effective temperature \(T_E(\alp)\) is shown in
\cref{fig:alp} for two random pixels in each zone.
Higher \alp implies a shorter penetration depth, reducing the thickness contributing to the signal. As the ice is colder near the surface than at depth (\cref{fig:scm}), \(T_E(\alp)\) is lower with higher \alp.
Following the same logic at the limits, \(\alp \rightarrow 0\) returns the bedrock temperature, while \(\alp \rightarrow \infty\) returns the surface temperature. The strong dependency of \(T_E(\alp)\) on \alp and the variety of dependencies among the different zones and pixels supports the underlying idea of inferring the absorption parameter \alp by comparing the spatial variations of observed and modeled \TB.

\begin{figure}[htb]
\centering
\subfloat{\includegraphics[width=\linewidth]{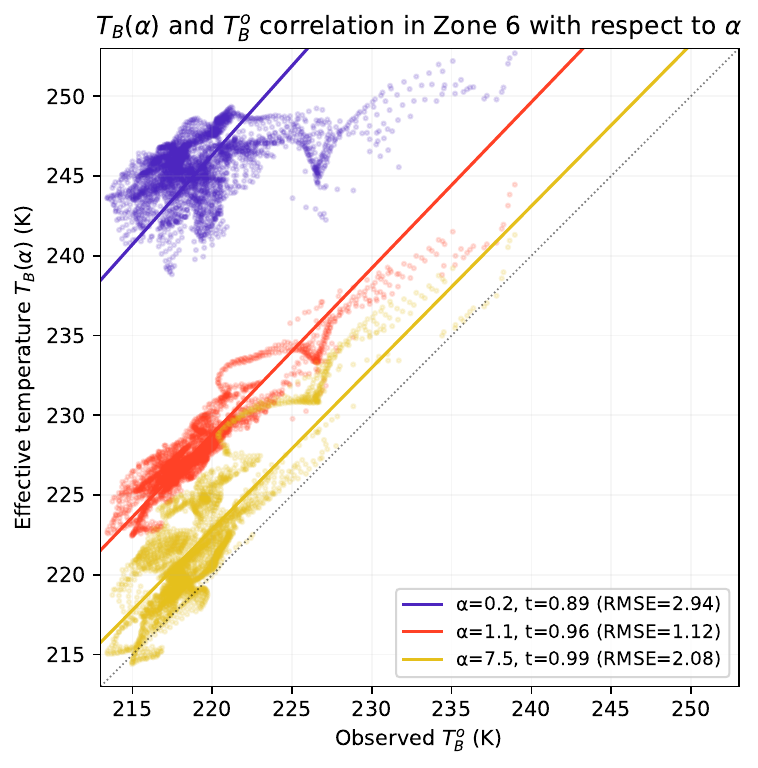}}%
\caption{Modeled effective temperature with respect to observed SMOS brightness temperatures for all points in \textcolor{zone6}{Zone~6} and a range of \alp. The line is the best fit for the transmittance \(\tra\). The \rmse of the fit is shown in the legend.}
  \label{fig:cld}
\end{figure}

\Cref{fig:cld} compares the modeled \(T_E(\alp)\) and observed \(\tb^o\) for three values of \alp in \textcolor{zone6}{Zone~6} along with the best fit for \(\tra\) and the resulting root mean squared error (\rmse).
Higher \alp result in lower modeled \(T_E\), consistent with \cref{fig:alp}.
Furthermore, the spread of the pixels also depends on \alp. 
Low \alp (0.2 in purple) clearly shows a large spread, denoting a poor model performance. 
Conversely, if all the dots were aligned with slope \(1/\tra\), the model would be perfect. The spread is visibly smaller for \(\alpha=1.1\) (red) than for higher (golden) or lower \alp, indicating a better performance of the model in that case which is reflected in the low RMSE\,=\,1.12\,K.

In \textcolor{zone6}{Zone~6}, the best linear relationship between \(T_E(\alp)\) and \(\tb^o\) is obtained for the value \(\alp = 1.11\) (red).
In all other zones, the optimal \alp is close, typically between 1 and 1.26.
This demonstrates the principle of our method, but to take into account the correlation between pixels and to evaluate the uncertainty associated with the resulting estimates, we further apply the Bayesian inference method depicted in \cref{sec:mem}.

\subsection{Resulting Posterior Distributions}
\label{ssc:pst}

\begin{figure}[hbt]
  \centering

  \includegraphics[width=\linewidth]{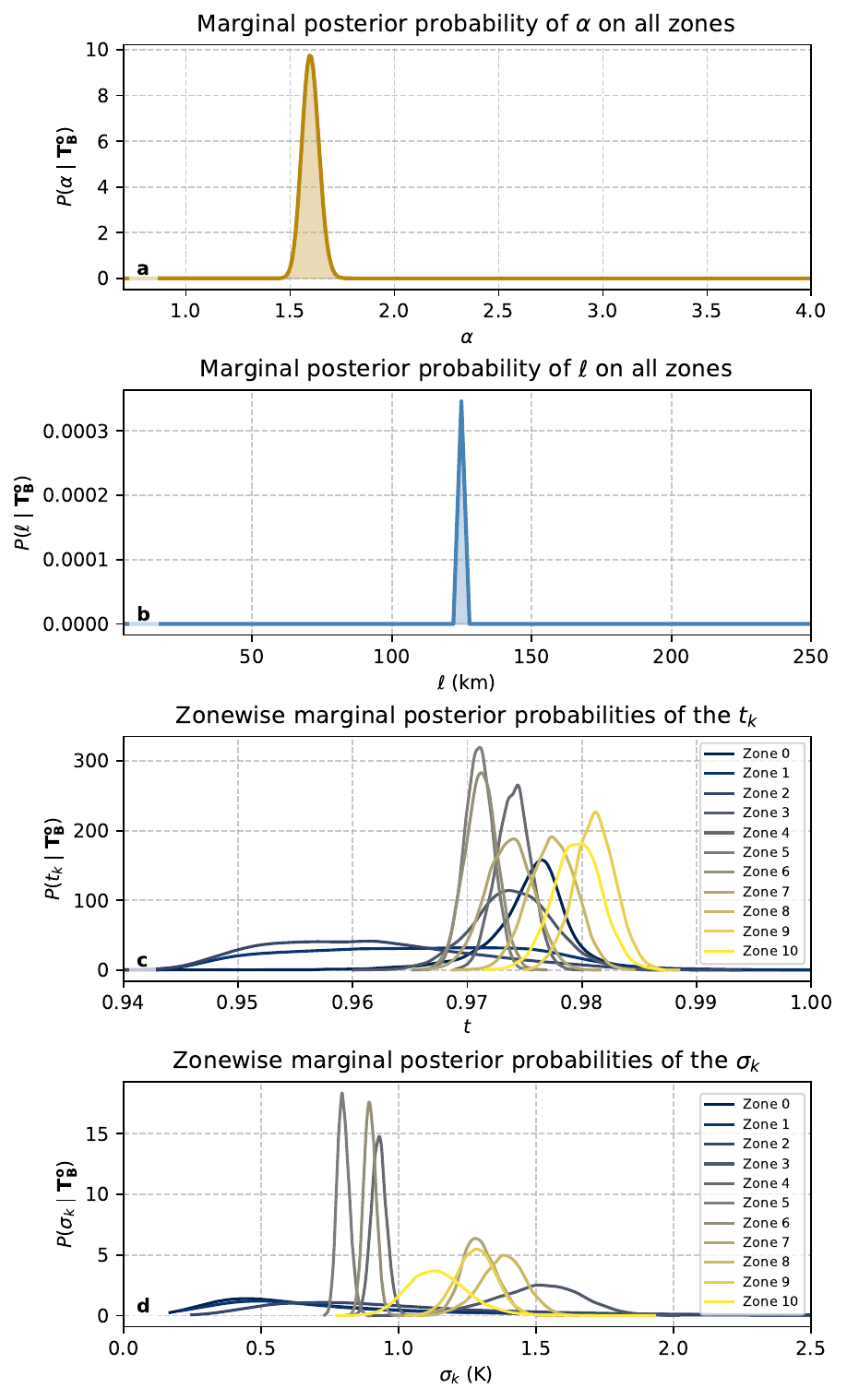}
  \caption{Marginalized posterior \pdf of \alp (a), \lda (b), the \(\tra_k\) (c) and the \(\sigma_k\) (d).}
  \label{fig:pst}
\end{figure}

The posterior distributions for \alp, \lda and the \(\tra_k\) and \(\sigma_k\) for each zone \(k\) are presented in \cref{fig:pst} as marginal distributions over each parameter.
Figure~\ref{fig:pst}a shows that the posterior distribution of \alp is peaked, thin and symmetric. 
The expectancy and maximum \textit{a posteriori} (\map) of \alp are very close, equal to 1.6.
The 99\% credible interval is between 1.5 and 1.7, resulting in an optimal absorption of \(\alpha = 1.6 \pm 0.1\), the main result of this study.

\Cref{fig:pst}b shows that the distribution of \lda is extremely narrow which means our method highly constrains the correlation between pixels. The resulting correlation length of 120~km, almost three times larger than the satellite footprint, confirms that this footprint is not the sole source of spatial correlation, also arising from model and geophysical parameter uncertainty. This large correlation length limits the model's validity in zones \textcolor{zone0}{0}-\textcolor{zone2}{2}. They contain less than 100 pixels each (other zones contain around 1000 pixels) and are in multiple locations thinner than the retrieved correlation length of \cref{fig:pst}b, making zone-specific results uninformative. Moreover, we show less confidence in the simplified emission model where the transmittance and polarization ratio are low. We therefore exclude zones \textcolor{zone0}{0}-\textcolor{zone2}{2} from the following analysis. 

\begin{figure}[htb]
  \includegraphics[width = \linewidth]{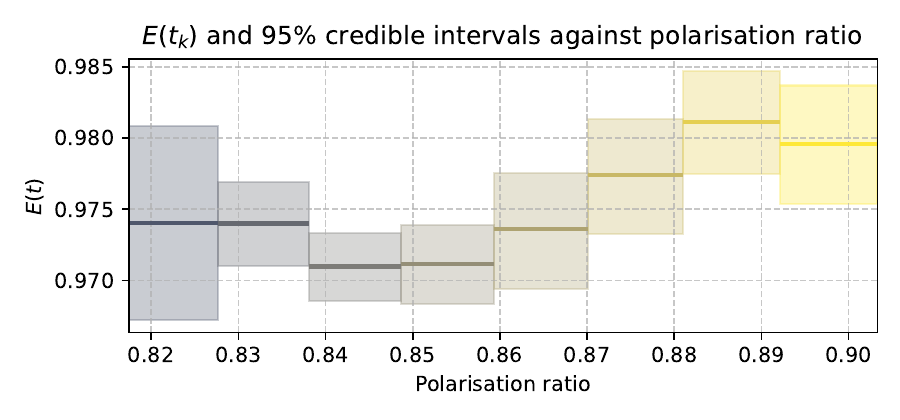}
  \caption{Mean of sampled \(\tra_k\) and 95\% confidence intervals with respect to Zone \(k\)'s polarization ratio range.}
  \label{fig:eta}
\end{figure}

The distributions of the \(\tra_k\) are less peaked and vary with \(k\) (Figure~\ref{fig:pst}c).
The estimated transmittance range \(0.96 < \tra < 0.99\), close to 1, shows a small contribution of the layering, in agreement with \cite{passalacqua2018}.
\Cref{fig:eta} directly shows the relationship between each zone's polarization ratios and the \(\tra_k\) posterior distributions. Where the polarization ratio increases, the transmittance of firn increases, with lower variations. Considering that \tra varies little in zones of bounded polarization ratios is confirmed as a valid hypothesis.

The credible values of the \(\sigma_k\) range between 0.7 and 1.8~K with no clear dependence on \(k\) (\cref{fig:pst}d). They are much higher than the observation error (a 3~K standard deviation for individual satellite observation, reduced by averaging over 10 years of sub-daily measurements \cite{kerr10}), suggesting that the source of the error \(\err\) is mainly the model or the geophysical parameters. 

\begin{figure}[th]
\includegraphics[width=\linewidth]{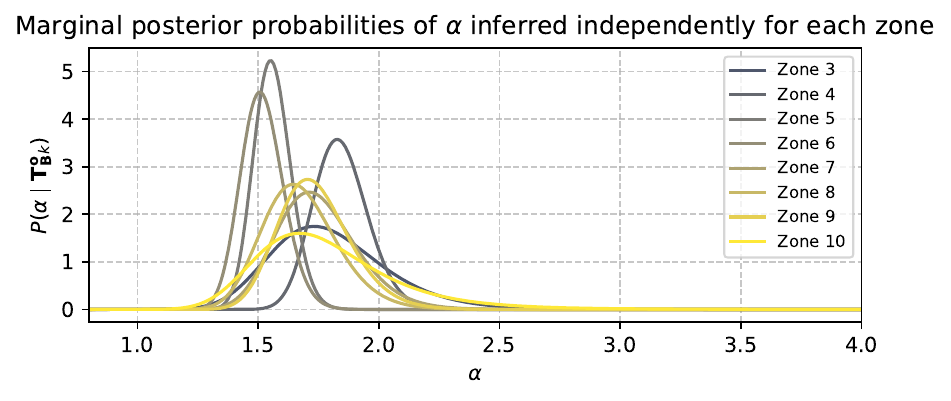}
\caption{Marginal \pdf of \(\alp\) computed for each zone \(k\) independently.}
\label{fig:mza}
\end{figure}

\Cref{fig:mza} shows the posterior distribution of \alp when applying the method independently in each zone. There is little difference in the most likely values for each zone, with \map ranging from 1.5 to 1.9.
The credible values of \alp contrast with the optimal values \(1\leq\alp\leq1.26\) found in \cref{ssc:vis}. The difference of approximately 0.4 between the \map and the optimal value of \(\alp\) found by minimizing the \rmse can be explained by the inclusion of correlation between pixels geographically close. Thin regions within each zone will be given too little confidence when computing only the \rmse~(equivalent of \(\Sigma = I\)) compared to wider regions, which can alter the resulting retrieved parameters.
By including spatial correlation, our framework allows for appropriate confidence in each observation, resulting in a more precise evaluation of credible \alp values.

\subsection{Hypotheses Evaluation}
\label{ssc:hyp}

\begin{figure}[hbt]
  \centering

  \subfloat[Expectancy of \(\alpha\) using \(\kao = \text{M06}\), computed over each zone independently against the zone's mean \TB.]
  {
    \includegraphics[width=\linewidth]{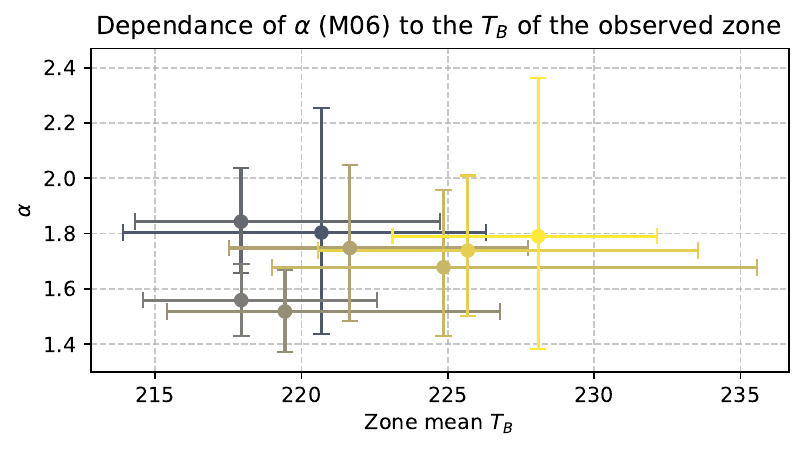}
      \label{fig:mae}
  }

  \subfloat[Expectancy of \(\alpha\) using \(\kao = \text{T84}\), computed over each zone independently against the zone's mean \TB.]{
    \includegraphics[width=\linewidth]{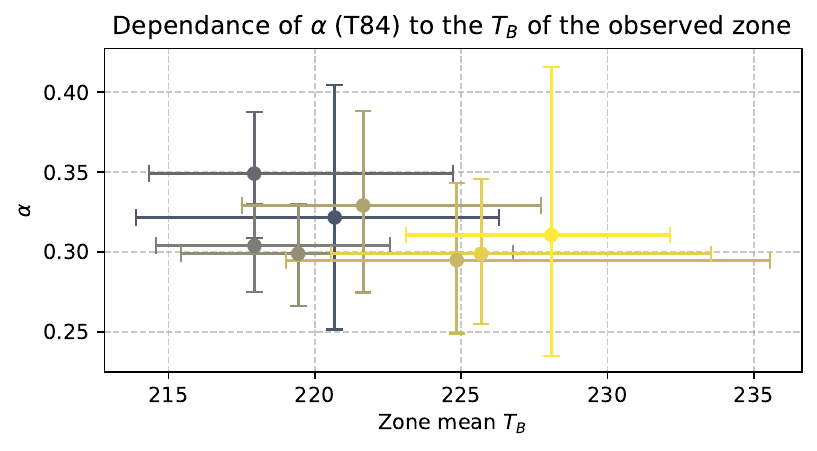}
      \label{fig:tbe}
  }
  \caption{Expectancy of \alp computed over Zones 3-10 independently against the zone's mean \TB. Vertical error bar represents the 90\% credible interval for the \pdf of \alp. Horizontal error bars show the statistical distribution of 90\% of the zone's \TB. Each color represents a zone (\cref{fig:map}).}
  \label{fig:tiu}
\end{figure}

Our retrieval framework builds on several assumptions which can be assessed \textit{a posteriori} from the results. 
Considering \(\ka(T) = \alpha\kao(T)\) assumes a correct temperature dependence of \kao, which can be verified by assessing the dependence of the retrieved \alp on the internal ice temperature for the pixels used in the retrieval.
The \(\alp\) retrieved with \(\kao = \text{M06}\) using a single zone are shown in \cref{fig:mae} with respect to the zone's mean \(\tb^o\).
The \alp retrieved with \(\kao = \text{T84}\), a formulation with a different dependence to ice temperature, are shown in \cref{fig:tbe}.
None show a clear dependence on temperature, so the assumption does not suffer a clear bias.
However, considering \(\ka(T) = \alpha\kao(T)\) does not allow to retrieve information on the temperature dependency even when using multiple formulations \kao.

\begin{figure}[htb]
    \centering
    \includegraphics[width=\linewidth]{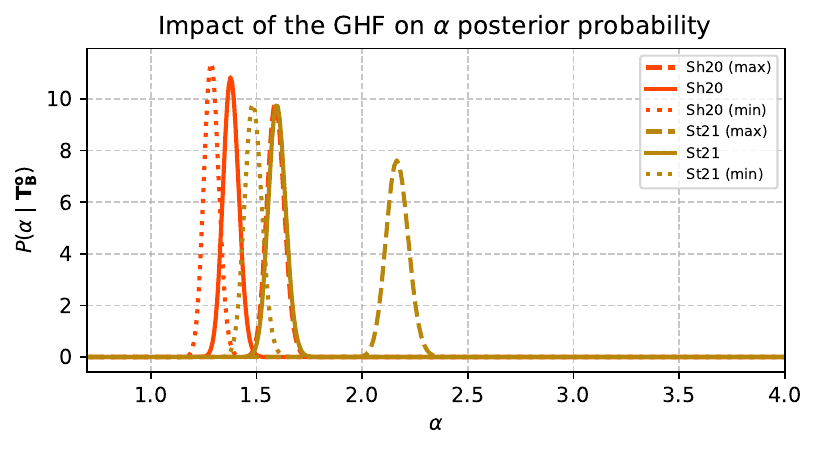}
    \caption{Marginal posterior probability of \alp using six geothermal heat flow maps: St21 from Stål \etal \cite{stal2021}, Sh20 from Shen \etal \cite{shen20} along with the minimal and maximal values given by their uncertainty bounds.}
    \label{fig:ghf}
\end{figure}

Among the input parameters, the geothermal heat flux is known to be the most uncertain. We tested the influence of the GHF on the retrieved absorption by computing the posterior probability of \alp using the boundary values from two widely used GHF datasets \cite{stal2021,shen20}.
The results are presented in \cref{fig:ghf}. 
The lack of knowledge about the GHF produces a larger uncertainty on \alp than the credible distribution of 1.5--1.7 retrieved with our method, with values ranging from 1.2 to 2.3 when testing the lower and upper boundaries of both datasets. 
While point-wise uncertainty on the GHF is included in the random \(\err\) error, a continent-wide bias clearly leads to a bias in the retrieved absorption. 

\begin{figure}[h!tb]
  \subfloat{\includegraphics[width=\linewidth]{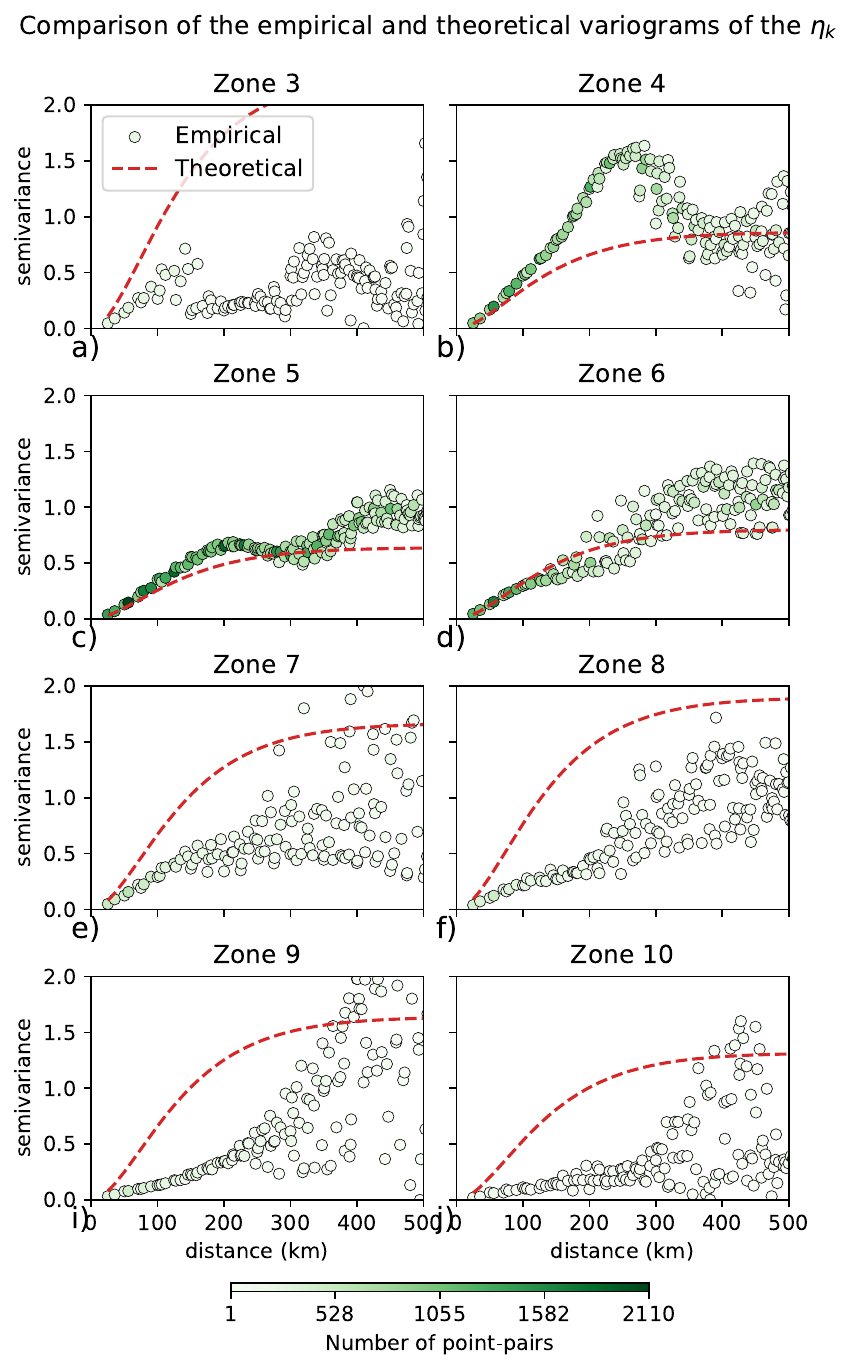}}
  \caption{Variogram of \(\tb^{o} - \tra_k^* T_E(\alpha^*)\) compared to the chosen covariance model (dashed, red) with inferred \(\sigma^*_k, \lda^*\) where \(^*\) indicates the \map. Each circle represents a distance, colored by the number of pairs of pixels at this distance in Zone \(k\).}
  \label{fig:var}
\end{figure}

As mentioned in \cref{sss:imp}, estimating \(\sigma\) and \lda simultaneously with a Matérn kernel is notoriously difficult. 
We evaluate the covariance model with a comparison of the theoretical and numerical variograms of the error \(\err\) (\cref{fig:var}).
The theoretical semi-variance fits the observed error precisely for Zones 4-6 and less for the other zones, which contain fewer pixels. Nonetheless, the Matérn 3/2 kernel was chosen as the one which fits the best the empirical variogram, compared to other Matérn kernels with different smoothness parameters and to Gaussian kernels. The latter are numerically close to singularity which forces approximations.

Finally, the loss of information due to the downsampling of observations was evaluated using each of the four nearby grids generated by sampling one every other 12.5 km pixel, starting from a different central pixel. We compare the \map for \alp for our retrieval and the three other resulting 25 km grids. Using these other grids results in the \map \(\alp^*\) ranging from 1.55 to 1.65, well within the credible interval of 1.5-1.7.

\section{Discussion}
\label{sec:dis}

\begin{figure}[ht]
  \includegraphics[width=\linewidth]{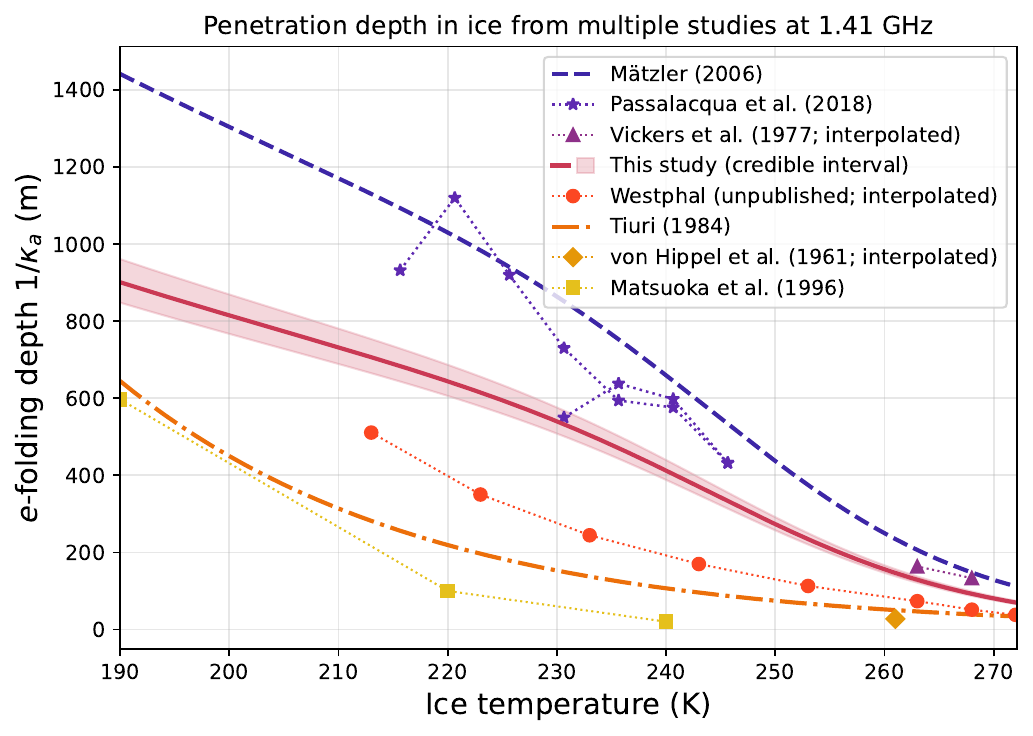}
  \caption{Penetration depth at 1.41 GHz according to multiple absorption formulations, following \cite{fujita00, warren84, matzler06, tiuri84, passalacqua2018}. When no formulation is given at 1.41 GHz is given, linear interpolation is used between surrounding frequencies. The \(e\)-folding depth \(1/\ka\) is the distance after which a transmitted signal is reduced by a factor \(e\), in ice of constant temperature.}
  \label{fig:abs}
\end{figure}

This study presents the first determination of a credible interval for absorption in the Antarctic ice sheet at 1.41 GHz, retrieved through Bayesian inference using the spatial variations of radiometric measurements from the SMOS satellite.
Our main result is an absorption value very likely higher than M06 \cite{matzler06} by a factor between 1.5 and 1.7. We further compare this result with several existing parameterizations of absorption, shown in \cref{fig:abs} as the e-folding depth ($1 / \ka$).
Our result falls in the  range of all the parameterizations and measurements. M06, P18 and Vickers \etal \cite{vickers75} have a larger penetration, while Westphal (unpublished, cited in \cite{warren84}),  T84, the compilation by Matsuoka \textit{et al} \cite{matsuoka96} and measurements by von Hippel \etal \cite{vonhippel71} penetrate less. Despite a careful modeling of uncertainties, our credible interval does not include any of the other sets of measurements. 
This variability among different estimates has two possible causes. Our estimate is representative of the absorption in the ice-sheet, where the ice contains impurities, notably sulfate from volcanoes \cite{gautier19}.
Furthermore, our methodology uses some assumptions that do not transcribe in our uncertainty model, namely the lack of bias in input parameters. As a consequence, we may underestimate the total uncertainties.

The difference between P18 and this study is more difficult to explain, as they both estimate absorption in the same ice-sheet region, and both use the same general principle. 
When first using uncorrelated variational inference (\cref{ssc:vis}), we obtained results closer to P18.
We show with our Bayesian method that a significant spatial correlation exists in the error between model and observations up to a few hundreds of kilometers, and it impacts the results, not only the uncertainties but also the final value.

The high spatial correlation in the errors is probably dominated by the internal temperature profile error correlation.
While any \(T_E\) bias is absorbed by the \tra variable, a consistent bias in internal temperature profiles may affect \alp retrieval, and this bias is not included in our estimated uncertainty.
To refine this point, future work should investigate internal temperature profile error, and devise a specific model for the propagation of this error.

In addition, the methodology we set up for retrieving ice absorption from internal temperature profiles and \TB measurements over ice sheets can be adapted to any new observations and geophysical parameter maps, other glaciological models and other parameterizations of the absorption. Any advance in the knowledge of internal temperature profiles or their uncertainty will greatly advance the study of microwave absorption in the ice sheets.
Our method uses a Bayesian framework with a limited number of variables, which enables high precision and fast computation of the posterior distribution. Future work may be dedicated to identifying variables to add to the framework in order to reduce the number of assumptions. 
In addition to variables related to internal temperature profile uncertainty, these may be other hyper-parameters on the probability distribution to weaken the constraints on the shape of the error distribution. Adding variables will necessitate advanced sampling or probability estimation methods to preserve the reasonable computation time and precision of the results. 

\section{Conclusion}

This study provides a new estimate of ice absorption at L-band under the real conditions of the Antarctic ice sheet. The main result is an absorption 1.6 times larger than the M06 formulation which is, to our knowledge, the lowest published estimate to date.
This suggests a 1.6 times shorter penetration of the microwaves through the ice with respect to this formulation. 
While several other studies such as T84 provided significantly larger absorption values, we found these to be incompatible with the amplitude of the \TB and surface temperature spatial variations in Antarctica. 

An important goal of this study was to estimate the uncertainties associated with the retrieval method. 
We found a credible interval of 1.6\(\pm\)0.1 for the multiplicative factor \alp on M06. The peaked distribution of \alp indicates a precise estimation of the absorption compared with the large spread among existing estimates in the literature.
We also found that an accurate modeling of the errors is critical, not only to provide precise uncertainties, but also as it influences the retrieved absorption value. This suggests further effort is necessary to model the different error sources. 
Another perspective is to apply our framework to lower frequencies as those observed by the UWBRAD spectral radiometer \cite{brogioni22}, as the absorption in the P-band is also fairly uncertain. This is necessary for the preparation of satellite missions exploring this low frequency range, such as CryoRad \cite{macelloni26}.

\appendices
\section{Marginalization}
\label{app:mar}
This section details the computations used to marginalize over \(\bm t\) and \(\bm\sigma\).
\begin{align*}
  &\p(\alpha, \lda \mid \bm{{\tb^{o}}}) 
  = \int \p( \alpha, \tra, \sigma, \lda \mid \bm{\tb^o}) d\sigma d t\\
  &\overset{\text{(Bayes)}}{\propto} \int \p(\alpha, \tra, \sigma, \lda)  \p( \bm{\tb^o} \mid \alpha, \tra, \sigma, \lda) d\sigma d t\\
  &\overset{\eqref{eqn:ind}}{\propto} \int \p(\alpha, \tra, \sigma, \lda)  \prod_{k=1}^K
  \frac{e^{-\frac12 \norm{\bm{T_{Bk}^o} - \tra_k\bm {T_{Bk}}(\alpha)}^2_{\inv{\paren{\sigma_k^2\Sigma_k(\lda)}}}}}{\sigma_k^{n_k}\sqrt{2\pi\left|\Sigma_k(\lda)\right|}}
  d\sigma d t\\
  &\overset{\text{(priors)}}{\propto} \p(\alpha, \lda) \int   \prod_{k=1}^K
  \frac{e^{-\frac12 \norm{\bm{T_{Bk}^o} - \tra_k\bm {T_{Bk}}(\alpha)}^2_{\inv{\paren{\sigma_k^2\Sigma_k(\lda)}}}}}{\sigma_k^{n_k+1}\sqrt{\left|\Sigma_k(\lda)\right|}}
  d\sigma d t\\
  &\overset{\text{(Fubini)}}{\propto} \p(\alpha, \lda)    \prod_{k=1}^K \int
  \frac{e^{-\frac12 \norm{\bm{T_{Bk}^o} - \tra_k\bm {T_{Bk}}(\alpha)}^2_{\inv{\paren{\sigma_k^2\Sigma_k(\lda)}}}}}{\sigma_k^{n_k+1}\sqrt{\left|\Sigma_k(\lda)\right|}}
  d\sigma_k d t_k\\
  &\overset{\eqref{tmp:1}}{\propto} \p(\alpha, \lda)    \prod_{k=1}^K \int
  \frac{\norm{\bm{T_{Bk}^o} - \tra_k\bm {T_{Bk}}(\alpha)}^{-n_k}_{\inv{\paren{\Sigma_k(\lda)}}}}{\sqrt{\left|\Sigma_k(\lda)\right|}}
   d t_k\\
   &\overset{\eqref{tmp:2}}{\propto} \p(\alpha, \lda)    \prod_{k=1}^K
  \frac{ \left(\min_t\norm{\bm{T_{Bk}^o}-t\bm{T_{Bk}}(\alpha)}_{\inv{\paren{\Sigma_k(\lda)}}} \right)^{-(n_k-1)} }{\norm{ \bm {T_{Bk}}(\alpha)}_{\inv{\paren{\Sigma_k(\lda)}}}\sqrt{\left|\Sigma_k(\lda)\right|}}
\end{align*}
where we used the formulas
\begin{equation}\label{tmp:1}
\int_{0}^\infty \frac{\exp(-A^2/\sigma^2)}{\sigma^{n+1}} d\sigma \propto A^{-n}
\end{equation}
for all $A>0$ and for all $u,v\in \mathbb R^n$,
\begin{align}
\int_{-\infty}^\infty \|u-tv\|_{\Sigma^{-1}}^{-n} d t
&\propto \frac{\left( \|u\|_{\Sigma^{-1}}^2-\tfrac{u^\top \Sigma v}{\|v\|_{\Sigma^{-1}}^2}\right)^{-\frac{n-1}{2}}}{\|v\|_{\Sigma^{-1}}} \nonumber\\
&=\frac{\left( \min_t\|u-tv\|_{\Sigma^{-1}}\right)^{-(n-1)}}{\|v\|_{\Sigma^{-1}}}. \label{tmp:2}
\end{align}

\section{Sampling}
\label{app:spl}
This section details the conditional probabilities of \(\bm\sigma\) and \(\bm t\).
For \(k\in \intset{0}{K-1}\) let \(a_k = 1/\norm{\bm{T_{Bk}}(\alp)}_{\inv{\Sigma_k(\lda)}}\), \(b = {\trsp{\bm{T_{Bk}}(\alp)}\inv{\Sigma_k(\lda)}\bm{T_{Bk}^o}}\) and \(c_k = \norm{\bm{T_{Bk}^o}}^2_{\inv{\Sigma_k(\lda)}}\).

\begin{align}
  \p(\sigma_k | \alpha,  \lda, \bm{\tb^o}) &= \frac{\p(  \alpha, \sigma_k, \lda | \bm{\tb^o})}{\p(\alpha,  \lda | \bm{\tb^o})}\\
  &\propto \frac{\frac{\p(\alpha, \lda)}{\sqrt{|\Sigma(\lda)|}} 
   \frac{\sqrt a_ke^{-\frac{c_k- a_k^2b_k^2}{2\sigma_k^2}}}{\sigma_k^{n}}}{p(\alpha, \lda)\sqrt{\frac{a_k}{|\Sigma(\lda)|}} \paren{c_k- a_k^2b_k^2}^{\frac{1 - n}{2}}}\\
   &\propto    \frac{\paren{c_k- a_k^2b_k^2}^{\frac{n-1}{2}}}{\paren{\sigma_k^2}^{\frac n2} }e^{-\frac{c_k- a_k^2b_k^2}{2\sigma_k^2}}
\end{align}
which shows
\begin{equation}
  \sigma_k^2 \mid \alpha,\lda,\bm{{\tb^{o}}} \sim \mathcal{IG}\left(\frac{n-1}{2},\ \frac{c_k-{b_k^2}{a_k^2}}{2}\right)
\end{equation}
where \(\mathcal{IG}\) is the inverse gamma distribution. In turn,
\begin{align}
  \p(\tra_k| \alpha,\sigma_k, \lda, \bm{\tb^o}) &= \frac{\p(  \alpha,\tra_k, \sigma_k, \lda | \bm{\tb^o})}{\p(\alpha,  \sigma_k, \lda | \bm{\tb^o})}\\
  &\propto \frac{\frac{\p(\alpha, \lda)}{\sqrt{|\Sigma(\lda)|}} 
  \frac{e^{-\frac12\paren{\frac{\paren{\tra_k -  a_k^2b_k}^2}{{\sigma_k^2 a_k}}+ \frac {c_k- a_k^2b_k^2}{\sigma_k^2}}}}{\sigma_k^{n+1}}}{\frac{p(\alpha, \lda)}{\sqrt{|\Sigma(\lda)|}} 
  \frac{\sqrt a_ke^{-\frac{c_k- a_k^2b_k^2}{2\sigma_k^2}}}{\sigma_k^{n}}}\\
  &\propto \frac{ 
  {e^{-\frac12\paren{\frac{\paren{\tra_k -  a_k^2b_k}^2}{{\sigma_k^2 a_k}}}}}}
  {\sigma_k
  {\sqrt a_k}}
\end{align}
results in
\begin{equation}
  \tra_k \mid \sigma_k,\alpha,\lda,\bm{{\tb^{o}}} \sim \mathcal N\left(a_k^2b_k,a_k{\sigma_k^2}\right)
\end{equation}
\section*{Acknowledgment}
This work was supported by the European Space Agency (ESA) through the CryoRad Earth Explorer 12 Phase 0 Science and Requirements Consolidation Study (SciReC, Grant 4000145903/24/NL/IB/ar).


\bibliographystyle{IEEEtran}
\bibliography{bibtex/bib/IEEEabrv,bibtex/bib/ref}

\ifCLASSOPTIONcaptionsoff
  \newpage
\fi

\begin{IEEEbiographynophoto}
{Hippolyte Signargout} received the M.Sc. degree in Operations Research from the University of Edinburgh in 2019 and the Ph.D. degree in computer science from the École normale supérieure de Lyon in 2023. He uses his background in theoretical computer science to develop efficient inverse methods for remote sensing, including Bayesian retrieval of ice-sheet characteristics from satellite observations.
\end{IEEEbiographynophoto}

\begin{IEEEbiographynophoto}
    {Ghislain Picard} received the M.Sc. degree in remote sensing from the University of Paris VII, Paris, France, in 1997, and the Ph.D. degree from the Center d’Etudes Spatiale de la Biosphere (CESBIO), Toulouse, France, in 2002.
In 2005, he joined the Institut des Géosciences de l’Environnement, Université Grenoble-Alpes, Grenoble, France, where since 2018, he has been a Full Professor. He is involved in the development of innovative instruments for the characterization of the snow-physical properties and of optical and microwave radiative transfer (RT) models. His research interests include the antarctic climate observed through the study of snow using a variety of remote sensing techniques and field experimentation.
\end{IEEEbiographynophoto}
\begin{IEEEbiographynophoto}
    {Olivier Zahm} is a researcher at Inria-Grenoble (France) whose work focuses on applied mathematics, uncertainty quantification, and computational methods for high-dimensional problems. His research develops efficient mathematical and statistical approaches for complex models, with applications in scientific computing and data analysis.
\end{IEEEbiographynophoto}

\begin{IEEEbiographynophoto}
{Ange Haddjeri} received the M.Sc. degree in oceanic, atmospheric, and climate science from Claude Bernard University Lyon 1, Villeurbanne, France, in 2020, and the Ph.D. degree in alpine snowpack variability modeling from Paul Sabatier University, Toulouse, France, in 2024.
His research interests include Alpine blowing snow variability, snow and ice modeling, and the application of visible and microwave remote sensing.
\end{IEEEbiographynophoto}
\begin{IEEEbiographynophoto}{Marion Leduc-Leballeur} received the M.Sc. degree in remote sensing and the Ph.D. degree in atmospheric physics from the University Pierre and Marie Curie (UPMC), Paris, France, in 2008 and 2012, respectively. From 2012 to 2016, she was with the Institute of Environmental Geosciences (IGE), Grenoble, France, focusing on using L-band observations from the Soil Moisture and Ocean Salinity (SMOS) satellite to investigate snow properties of the Antarctic ice sheet. In 2017, she joined IFAC-CNR, Sesto Fiorentino, Italy, as a Research Scientist, working on the study of snow and ice properties from in situ measurements, satellite observations, and electromagnetic models. Her main field of research is the study of the cryosphere using passive microwave satellite observations.
\end{IEEEbiographynophoto}

\begin{IEEEbiographynophoto}
{Pierre Zeiger} received the Ph.D. degree in continental hydrology from Paul Sabatier University, Toulouse, France, in 2022. Since then, he is working as a post-doctoral researcher in snow physics and remote sensing at the Institut des Géosciences de l'Environnement (IGE), Grenoble, France. His research interests include remote sensing of polar snow, firn hydrology and links with ice sheets mass loss.
\end{IEEEbiographynophoto}

\begin{IEEEbiographynophoto}{Marco Brogioni} received the M.Sc. degree in telecommunications engineering with the University of Siena, Italy, in 2003, and the Ph.D. degree in remote sensing with the University of Pisa, Italy, in 2008. Since 2004, he has been with the Microwave Remote Sensing Group, Institute of Applied Physics "Nello Carrara" (IFAC), National Research Council of Italy (CNR), Florence, Italy. From 2006 to 2007, he has been a Visiting Student at the University of California at Santa Barbara, Santa Barbara, CA, USA working on microwave modelling of snow covered surfaces. His research interests include passive and active microwave remote sensing applied to snow by using satellite and ground-based data, especially development of electromagnetic models for passive and active microwave remote sensing of snow, vegetation, and soil. Most of his activities are carried out in the framework of ESA and ASI projects. He is also involved in the design and manufacturing of microwave radiometers (P- to Ka-bands). Since 2021 leads the development of the Cryorad ultrawideband airborne demonstrator in the framework of an Italian Space Agency project. He is currently involved in several international projects regarding polar regions. He is part of ESA SMOS L1 team and former but still active NASA ESTO UWBRAD team. He was the recipient of the Third Prize at the URSI GA Student Prize Paper Competition in Chicago, IL, USA, in 2008. He served as co-chair of the 16th MicroRad 2020 IEEE meeting and in the local organizing committee of the 10th Microrad, Florence, in 2008 and the Microwave Signature Symposium of the URSI Commission-F, Florence, in 2010. He participated the Italian Antarctic Expeditions as PI and field campaign leader, carrying out his research at Concordia Station (Dome-C) and Mario Zucchelli Station (Ross Sea) in 2013, 2015 and 2018. At present he is the PI of the ESA Earth Explorer 12 CryoRad Phase 0 and Phase A scientific study.
\end{IEEEbiographynophoto}

\begin{IEEEbiographynophoto}
    {Giovanni Macelloni} (Senior Member, IEEE) received the M.Sc. degree in electronic engineering from the University of Florence, Florence, Italy, in 1993.
He is currently the Director of the Institute of Applied Physics “Nello Carrara”—National Research Country, Sesto Fiorentino, Italy, and an Adjunct Professor with the University of Venice. He is also the Science Lead of the ESA’s EE12 Candidate Mission CryoRad. He is the National Delegate of SCAR and Global Cryosphere Watch of the WMO. He has worked as a tutor for Ph.D. students. His research has been carried out in the framework of several national and international programs granted by Italian entities, European Community, and Space agencies (ESA, ASI, NASA, and JAXA), and includes participation in international teams for the study of the cryosphere and the development and assessment of future spaceborne missions. His research interests include microwave active and passive remote sensing for the study of the Earth system and the cryosphere in particular. He is also involved in the design and development of microwave remote sensing systems, both from the ground, the air, and satellites.
Mr. Macelloni served as a reviewer for several international committees, organizations, and international journals, and acted as an organizer and the co-chair of international conferences. He is an Associate Editor of IEEE Transactions on Geoscience and Remote Sensing.
\end{IEEEbiographynophoto}

%



\end{document}